# Graphene-Supported Au-Ni Carbon Nitride Electrocatalysts for the ORR in Alkaline Environment

E. Negro[a,b], A. Bach Delpeuch[c], K. Vezzù[c], S. Polizzi[d], F. Bertasi[c], G. Nawn[c], G. Pagot[b,c], V. Di Noto[c,e*]

[a] Department of Chemical Sciences, University of Padova, Via Marzolo 1, I-35131 Padova (PD), Italy

[b] Centro Studi di Economia e Tecnica dell'Energia *"Giorgio Levi Cases"*, Via Marzolo 9, I-35131 Padova (PD), Italy

[c] Department of Industrial Engineering, University of Padova, Via Marzolo 1, I-35131 Padova (PD), Italy, in the Department of Chemical Sciences

[d] Department of Molecular Sciences and Nanosystems and Centre for Electron Microscopy *"G. Stevanato"*, University *"Ca' Foscari"* of Venice, Via Torino 155/B, I-30172 Venezia-Mestre (VE), Italy

[e] Consorzio Interuniversitario Nazionale per la Scienza e la Tecnologia dei Materiali - Italy

*Corresponding author. E-mail address: vito.dinoto@unipd.it

This study reports the preparation and characterization of a new family of electrocatalysts (ECs) for the oxygen reduction reaction (ORR) exhibiting a *"core-shell"* morphology. The *"core"* consists of graphene sheets, which are covered by a carbon nitride (CN) *"shell"* embedding Au and Ni active sites. The investigated ECs are labeled AuNi$_{10}$-CN$_1$ 600/Gr and AuNi$_{10}$-CN$_1$ 900/Gr. The chemical composition and thermal stability are studied by inductively-coupled plasma atomic emission spectroscopy (ICP-AES), elemental analysis and by high-resolution thermogravimetric analysis (HR-TGA). The morphology of the ECs is probed by scanning electron microscopy (SEM), high-resolution transmission electron microscopy (HR-TEM) and powder X-ray diffraction (XRD). The ORR performance of the ECs is studied both in acid (0.1 M HClO$_4$) and in alkaline medium (0.1 M KOH) by Cyclic Voltammetry with the Thin-Film Rotating Ring-Disk Electrode (CV-TF-RRDE) method. Both ECs exhibit a promising performance in the ORR in the alkaline medium.

**Introduction**

The oxygen reduction reaction (ORR) is one of the most widely studied processes in modern electrochemistry, and is involved in the operation of several families of advanced electrochemical energy conversion/storage devices (*e.g.*, fuel cells, FCs, and metal-air batteries). The ORR exhibits a complex mechanism, which is strongly affected by the environmental conditions (*e.g.*, pH, nature of the solvent) and typically involves a number of different steps (*e.g.*, electron transfer, bond breaking, bond formation) (1-4). For these reasons, the kinetics of the ORR are very sluggish, leading to large activation overpotentials. On these bases, the development of electrocatalysts (ECs) able to promote effectively the ORR in FCs and metal-air batteries, and consequently raise the energy conversion efficiency of the device to match the requirements set by practical applications is a major goal of the research.

Typical ORR ECs for practical electrochemical energy conversion/storage devices must operate in very harsh environments in terms of: (i) high current densities; (ii) strongly oxidizing conditions; and (iii) extreme pH values (*i.e.*, strongly acid for proton-exchange membrane fuel cells, PEMFCs, and highly alkaline for metal-air batteries and anion-exchange membrane fuel cells, AEMFCs). The ECs must also exhibit a long operating lifetime which, for some applications (*e.g.*, stationary PEMFCs) must reach a few tens of thousands of hours. In order to achieve these demanding requirements one of the best approaches is to develop nanocomposite ORR ECs, comprising: (i) mono/plurimetallic nanoparticles (NPs), which are supported on (ii) a suitable support exhibiting a large surface area and a high electrical conductivity. In the case of ORR ECs for the acid environment, the NPs must typically be based on platinum-group metals (PGMs) to achieve a sufficient performance/durability level. On the other hand, if the ORR is carried out in the alkaline environment, other elements can be taken into consideration (*e.g.*, Au, Ag and Ni), significantly mitigating the risk of supply bottlenecks. Typical supports for ORR ECs include carbon blacks such as Vulcan XC-72R, which exhibits a very high electrical conductivity and large surface area, on the order of *ca.* 250 $m^2·g^{-1}$ (5). Finally, it is highlighted that the interface between the metal NPs bearing the ORR active sites and the support is very sensitive to degradation effects, which are known to reduce the durability of the system. The latter include: (i) NP detachment or coalescence, which reduce the active surface area of the EC; and (ii) oxidation of the support, hindering the electron transfer between the active sites and the external circuit.

On these bases, in order to improve the performance and durability of ORR ECs, it is necessary to develop new materials exhibiting the following features:

- *Faster ORR kinetics.* This objective can be achieved by modulating the chemical composition of the metal NPs. A typical approach consists in alloying the *"active"* metal (*e.g.*, a PGM, Au, or Ag) with a first-row transition metal (*e.g.*, Co, Ni). Very good results are obtained by devising NPs with a *"skeleton-skin"* structure, where an alloyed *"skeleton"* (*e.g.*, comprising Pd and Co) is covered by a thin *"skin"* of the pristine *"active"* metal (*e.g.*, Pd) (6).
- *An improved support.* One possibility is to use graphene sheets as the support (7, 8). Indeed, the latter exhibit: (i) a very high electrical conductivity (minimizing the ohmic drops); (ii) a large specific surface area, up to *ca.* 2600 $m^2 \cdot g^{-1}$; and (iii) a low microporosity (which facilitates the mass transport of reactants and products).
- *A more stable interface between the metal NPs and the support.* In this regard, a solution is to devise ECs characterized by a *"core-shell"* morphology; the support *"core"* is covered homogeneously by a thin graphitic carbon nitride (CN) *"shell"* coordinating the metal NPs in C- and N- *"coordination nests"*. The concentration of N in the CN *"shell"* is low (lower than *ca.* 5 wt%), to achieve a facile electron transport from the active sites to the external circuit (9, 10).

The ORR ECs described here include: (i) graphene sheets as the *"core"*, covered by (ii) a graphtic CN *"shell"*, embedding (iii) metal NPs including Au as the *"active metal"* and Ni as the *"co-catalyst"*. Two ORR ECs are described herein: *"AuNi$_{10}$-CN$_l$ 600/Gr"* and *"AuNi$_{10}$-CN$_l$ 900/Gr"*, which are labelled in accordance with the nomenclature proposed elsewhere (11), and are distinguished on the basis of $T_f$ (*i.e.*, the temperature of the final step of the pyrolysis process). $T_f$ = 600 and 900°C respectively for AuNi$_{10}$-CN$_l$ 600/Gr and AuNi$_{10}$-CN$_l$ 900/Gr. The proposed ECs are optimized for the alkaline environment, hence the choice to adopt Au as the *"active metal"*. The nominal Au : Ni molar ratio in the ECs is equal to 1 : 10. This value is adopted to explore the feasibility to obtain a system characterized by a hierarchical morphology, where Ni acts a *"sacrifical component"* and is removed by electrochemical dealloying before the determination of the ORR performance by Cyclic Voltammetry with the Thin-Film Rotating Ring-Disk Electrode (CV-TF-RRDE) setup, similarly to an approach already described in the technical literature (12).

## Experimental

Reagents

Potassium tetrachloroaurate (98 wt%), potassium tetracyanonickelate (II), hydrate, (D$^+$) sucrose, biochemical grade and graphene nanoplatelets are obtained respectively from Abcr GmbH, Sigma, Acros Organics and ACS Material. Hydrofluoric acid (48 wt.%), perchloric acid (67-72 wt%) and potassium hydroxide (98.4%) are supplied by Sigma-Aldrich, Fluka Analytical and VWR International, respectively. Isopropyl alcohol (purity > 99.8 wt%) is acquired by Sigma-Aldrich. A commercial Nafion 1100 solution (5 wt%) is procured from Alfa Aesar. XC-72R carbon black (provided by Carbocrom s.r.l as a courtesy) is treated in H$_2$O$_2$ (10 vol. %) prior to the experiments. A commercial 10 wt.% Pt/C electrocatalyst (ElectroChem, Inc.), labeled in this manuscript *"Pt/C ref."*, is employed as the reference material for the electrochemical analysis. All the chemicals are used as received, without any further purification treatment. Doubly distilled water is employed in all the experiments.

Synthesis of the Electrocatalysts (ECs)

The preparation of the ECs is carried out in accordance with the following protocol, inspired by a procedure described elsewhere (12). Briefly, 780 mg of sucrose is dissolved into the minimum amount of water; the resulting solution is divided into two aliquots, A and B. 1194 mg potassium tetracyanonickelate (II), hydrate are dissolved into A; subsequently, 390 mg of graphene nanoplatelets are further added and the resulting suspension is thoroughly homogenized by probe sonication. B is prepared exactly as A, with the difference that no potassium tetracyanonickelate is adopted; 131 mg of potassium tetrachloroaurate are added instead. B is mixed to A and the resulting product is thoroughly homogenized by probe sonication. The products undergo the typical pyrolysis and post-pyrolysis steps described in the literature (12, 13). Two ECs are finally obtained, labelled *"AuNi$_{10}$-CN$_l$ 600/Gr"* and *"AuNi$_{10}$-CN$_l$ 900/Gr"*. The final step of the pyrolysis process T$_f$ adopted to obtain AuNi$_{10}$-CN$_l$ 600/Gr and AuNi$_{10}$-CN$_l$ 900/Gr is carried out at 600°C and 900°C, respectively.

Instruments and methods

ICP-AES analysis is executed using a Spectro Arcos spectrometer equipped with an EndOnPlasma torch. The evaluation of the C, H, N and S wt% in the ECs is achieved by elemental analysis using a FISONS EA-1108 CHNS-O apparatus. HR-TGA investigation is conducted in the temperature range between 30 and 900°C with a heating rate changing as a function of the first derivative of the weight loss from 50°C/sec to 0.001°C/sec. A TGA 2950 analyzer (TA instruments) is employed to this aim. SEM images are collected by means of a Philips XL30TMP environmental scanning electron microscope (ESEM) at an acceleration voltage of 20 kV. Elemental X-ray fluorescent microanalyses were performed with an embedded EDX system coupled with an energy-dispersive X-ray spectrometer, equipped with a Si/Li detector. The HR-TEM images are recorded using a Jeol 3010 instrument (with a 0.17 nm point-to-point resolution) fitted with a Gatan slowscan 794 CCD camera. For that purpose, one drop of an isopropanol suspension containing the investigated EC is deposited on a holey 3 mm copper grid. Powder XRD patterns are acquired at 10° < 2θ < 90° with a 0.05° step, and a 30 sec integration time on an eXplorer diffractometer manufactured by GNR and mounting a monochromatized $CuK_\alpha$ source. The Rietveld analysis of the XRD spectra is executed with the MAUD software (Version 2.55).

Electrochemical characterization

A homogeneous blend composed of the EC and XC-72R carbon black (1:1 weight ratio) is prepared by grinding together the powders in an agate mortar. 25 mg of this blend is then weighted and put inside a vial where a 1 mL isopropanol/water mixture (1:1 v/v ratio) and 10 μL Nafion solution are added. The suspension is then homogenized by sonication. The Pt/C ref. is directly put into the vial without further addition of XC-72R carbon black. A measured aliquot of each suspension is then pipetted onto the glassy carbon disk (Ø 5 mm) of the RRDE tip; the solvent is removed and the loading of *"active metal"* (either Au, for the proposed ECs, or Pt, for the Pt/C ref.) is set at 15 µg·cm$^{-2}$. The Pt ring collection efficiency of the RRDE tip is equal to 0.38. The counter electrode is a Pt wire. The electrochemical measurements in acidic and alkaline medium are run using an Hg/Hg$_2$SO$_4$ and Hg/HgO reference electrode, respectively; the potentials are referred to the reversible hydrogen electrode (RHE). The data are collected on a multi-channel VSP potentiostat/galvanostat (Bio-Logic). The electrochemical performance of the ECs for the oxygen reduction reaction (ORR) is evaluated by cyclic voltammetry with the thin-film rotating ring-disk electrode (CV-TF-RRDE) setup in a three-electrode Teflon cell. The measurements are carried out at T = 25°C in O$_2$-saturated acidic (0.1 M HClO$_4$) and alkaline (0.1 M KOH) solutions. The working electrode is rotated at Ω = 1600 rpm. The ECs are activated by voltammetric cycling at ν = 100 mV·sec$^{-1}$ between 0.05 and 1.05 V *vs.* RHE in 0.1 M HClO$_4$ until the voltammogramms are stable. The electrolyte solution is then removed. The final measurements are collected after a further stabilization of the electrochemical activity of the electrocatalysts is guaranteed by potentiodynamic cycling at ν = 100 mV·sec$^{-1}$ between 0.05 and 1.05 V *vs.* RHE; the electrocatalytic performance is finally recorded during a CV at ν = 20 mV·sec$^{-1}$. The electrochemical results shown hereafter are obtained after elimination of the capacitive contribution accomplished by subtraction of the CVs run in an O$_2$-saturated electrolyte by those carried out in a N$_2$-saturated electrolyte as detailed elsewhere (13).

## Results and Discussion

The chemical composition of AuNi$_{10}$-CN$_l$ 600/Gr and AuNi$_{10}$-CN$_l$ 900/Gr is detailed in Table I. The pyrolysis process mostly affects the light elements such as potassium, hydrogen and nitrogen; their contents decrease as T$_f$ is raised. The lower amount of these elements at T$_f$ = 900°C indicates an improved graphitization of the CN matrix, which better coordinates Au and Ni (14). The molar ratio between Au and Ni closely matches the nominal values, and does not vary significantly on T$_f$, thus witnessing the potential of the proposed synthetic route to obtain ECs with a well-controlled chemical composition.

**TABLE I.** Chemical composition of the two CN ECs.

| Electrocatalyst | Atomic weight / wt.% | | | | | | Formula |
|---|---|---|---|---|---|---|---|
| | K$^{(a)}$ | Au$^{(a)}$ | Ni$^{(a)}$ | C$^{(b)}$ | H$^{(b)}$ | N$^{(b)}$ | |
| *AuNi$_{10}$-CN$_l$ 600/Gr* | 0.19 | 6.90 | 26.3 | 60.2 | 0.54 | 4.84 | K$_{0.14}$[AuNi$_{12.8}$C$_{143}$H$_{15.4}$N$_{9.86}$] |
| *AuNi$_{10}$-CN$_l$ 900/Gr* | 0.04 | 6.63 | 29.1 | 62.4 | 0.12 | 0.92 | K$_{0.03}$[AuNi$_{14.8}$C$_{154}$H$_{3.54}$N$_{1.94}$] |

$^{(a)}$ determined by ICP-AES.

$^{(b)}$ evaluated by elemental analysis.

Figure 1 shows the HR-TGA analysis for AuNi$_{10}$-CN$_l$ 600/Gr, AuNi$_{10}$-CN$_l$ 900/Gr and the Pt/C ref. performed in an oxidizing atmosphere. The HR-TGA profiles of the *"core-shell"* CN ECs exhibit two main events, I and II. I is evidenced at *ca*. 420°C and 470°C for AuNi$_{10}$-CN$_l$ 600/Gr and AuNi$_{10}$-CN$_l$ 900/Gr, respectively; it corresponds to the degradation of the CN *"shell"*. The higher decomposition temperature exhibited by AuNi$_{10}$-CN$_l$ 900/Gr is ascribed to a higher graphitization degree of its CN shell (5). II is revealed between 420°C and 650°C for AuNi$_{10}$-CN$_l$ 600/Gr and between 470°C and 720°C for AuNi$_{10}$-CN$_l$ 900/Gr; it is attributed to the combustion of the graphene support *"core"*. The high-temperature residue is ascribed to the metal atoms present in the ECs; its value matches well with the results obtained from the ICP-AES analysis and reported in Table I.

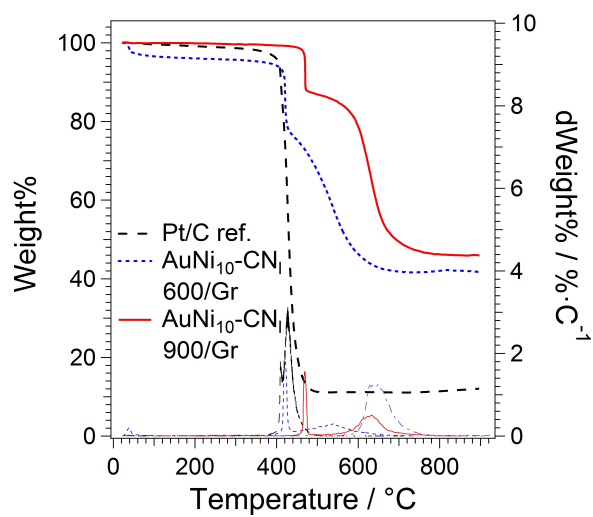

Figure 1. HR-TGA profiles in oxidizing atmosphere of the ECs.

The powder XRD patterns of AuNi$_{10}$-CN$_l$ 600/Gr and AuNi$_{10}$-CN$_l$ 900/Gr are depicted on Figure 2. A semiquantitative Rietveld analysis is performed to investigate structural features such as crystal phases included in the ECs, relative phase abundance and size of the nanoparticles (see Table II) (15). The Rietveld results highlight the presence of three different metallic phases for AuNi$_{10}$-CN$_l$ 600/Gr; one corresponds to pristine Ni, while the remaining two are associated to two different AuNi$_x$ systems: one Au-rich alloy, and a well-defined AuNi$_3$ intermetallic phase. These results indicate that at T$_f$ = 600°C the alloying between Au and Ni is incomplete. On the contrary, the XRD pattern of AuNi$_{10}$-CN$_l$ 900/Gr does not exhibit peaks which can be associated to pristine Ni. Two types of AuNi$_x$ fcc NP alloys are revealed; the former exhibits a low abundance (~1.5 wt%) and is characterized by a large Au:Ni molar ratio (*i.e.*, ~9), while the latter is largely dominant (~32 wt%) and exhibits a much smaller Au:Ni molar ratio (*i.e.*, ~0.08). Both ECs also include peaks (at 2θ ~ 26.6°) associated to graphene sheets which are still partially stacked. The thickness of these latter stacks decreases as T$_f$ is raised (AuNi$_{10}$-CN$_l$ 600/Gr → 35.5 nm; AuNi$_{10}$-CN$_l$ 900/Gr → 26.5 nm). This indicates that the pyrolysis process promotes the exfoliation of the graphene nanoplatelets used as the support.

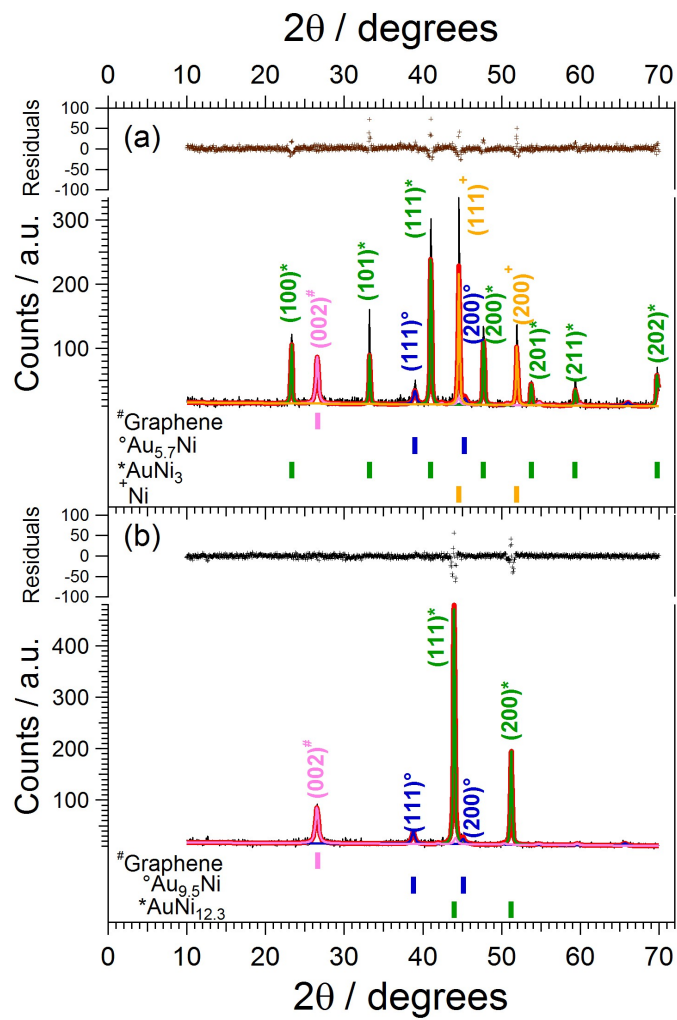

Figure 2. Powder XRD spectra with fitted Rietveld profiles of AuNi$_{10}$-CN$_l$ 600/Gr (a) and AuNi$_{10}$-CN$_l$ 900/Gr (b) are shown. The residues show the goodness of the fitting.

**Table II.** Structural information obtained from the Rietveld fit of the XRD patterns of Figure 2.

| | | AuNi$_{10}$-CN$_l$ 600/Gr | | |
|---|---|---|---|---|
| **Phase** | **Space group** | **Phase abundance / wt.%** | **Cell parameters / Å** | **Particle size / nm** |
| *Au$_{5.7}$Ni* | Fm-3m | 1.4 | a = 4.000 | 26.5 |
| *AuNi$_3$* | Fm-3m | 11.5 | a = 3.810 | > 90 |
| *Ni* | Fm-3m | 15.5 | a = 3.519 | > 90 |
| *Graphene* | - | - | c = 6.701 | 35.5 |
| | | AuNi$_{10}$-CN$_l$ 900/Gr | | |
| **Phase** | **Space group** | **Phase abundance / wt.%** | **Cell parameters / Å** | **Particle size / nm** |
| *Au$_{9.5}$Ni* | Fm-3m | 1.5 | a = 4.020 | 26.5 |
| *AuNi$_{12.3}$* | Fm-3m | 31.9 | a = 3.565 | > 90 |
| *Graphene* | - | - | c = 6.709 | 26.5 |

The SEM measurements (see Figure 3) show that the ECs proposed here present a morphology in accordance with similar materials (16). Indeed, the micrographs show micrometic / submicrometric bright grains, corresponding to the metal phases, dispersed through a dark background. The latter corresponds to the graphene *"cores"* covered by the CN *"shell"*. The size of the bright grains increases as T$_f$ is raised, in accordance with a typical mechanism of diffusion-controlled thermal growth (17). The chemical composition of the bright grains is inspected by energy-dispersive X-ray spectroscopy (EDX). The results match the outcome of the Rietveld analysis of the XRD patterns. In the case of AuNi$_{10}$-CN$_l$ 600/Gr (see Figure 3a), at least two distinct metallic phases are evidenced. The latter are identified with the Ni and AuNi$_3$ NPs previously revealed by

Rietveld analysis (see Figure 2), and further confirm that at $T_f = 600$ the alloying between Au and Ni is incomplete. On the other hand, EDX confirms that $AuNi_{10}$-$CN_l$ 900/Gr (see Figure 3b) comprises for the most part only one type of Au-Ni alloy NPs, with an approximate 1:10 molar ratio.

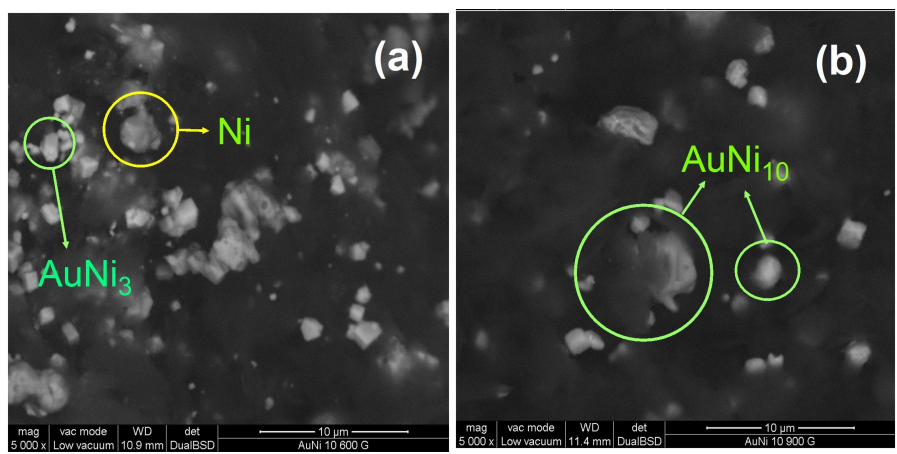

Figure 3. SEM images of $AuNi_{10}$-$CN_l$ 600/Gr (a) and $AuNi_{10}$-$CN_l$ 900/Gr (b) ECs.

The HR-TEM images of the *"core-shell"* ECs are shown in Figure 4. The low-resolution pictures (Figure 4a and Figure 4b) confirm that the metallic nanoparticles grow as the $T_f$ is raised from 600°C to 900°C (18). The CN *"shell"* embedding the metallic NPs, which is visible at medium resolution in Figure 4d, presents a thickness ranging from 20 to 100 nm. At high resolution (Figure 4e and Figure 4f) it is possible to evidence ordered planes, whose interplanar distance is determined for both ECs (see Figure 4g and Figure 4h). Figure 4g corresponds to an ordered domain exhibiting an interplanar distance of *ca.* 3.6 Å, which is attributed to the $d_{002}$ planes of the graphene *"core"* support of $AuNi_{10}$-$CN_l$ 600/Gr. Figure 4h, referring to $AuNi_{10}$-$CN_l$ 900/Gr, evidences NPs with interplanar distance equal to $d = 2.3$ Å. The latter is consistent with the $d_{111}$ planes of an $Au_{9.5}Ni$ alloy phase. The structural features of the nickel-rich $AuNi_{12.3}$ phase are difficult to be clearly identified by HR-TEM despite its very high abundance (see Table II). This result is interpreted admitting that in $AuNi_{10}$-$CN_1$ 900/Gr the alloy particles themselves exhibit a *"core-shell"* structure, which consists of: (i) an $AuNi_{12.35}$ *"core"*; the morphology of these NPs cannot be directly identified by HR-TEM owing to their large thickness, which hides the interplanar fringes; and (ii) an $Au_{9.5}Ni$ *"shell"*.

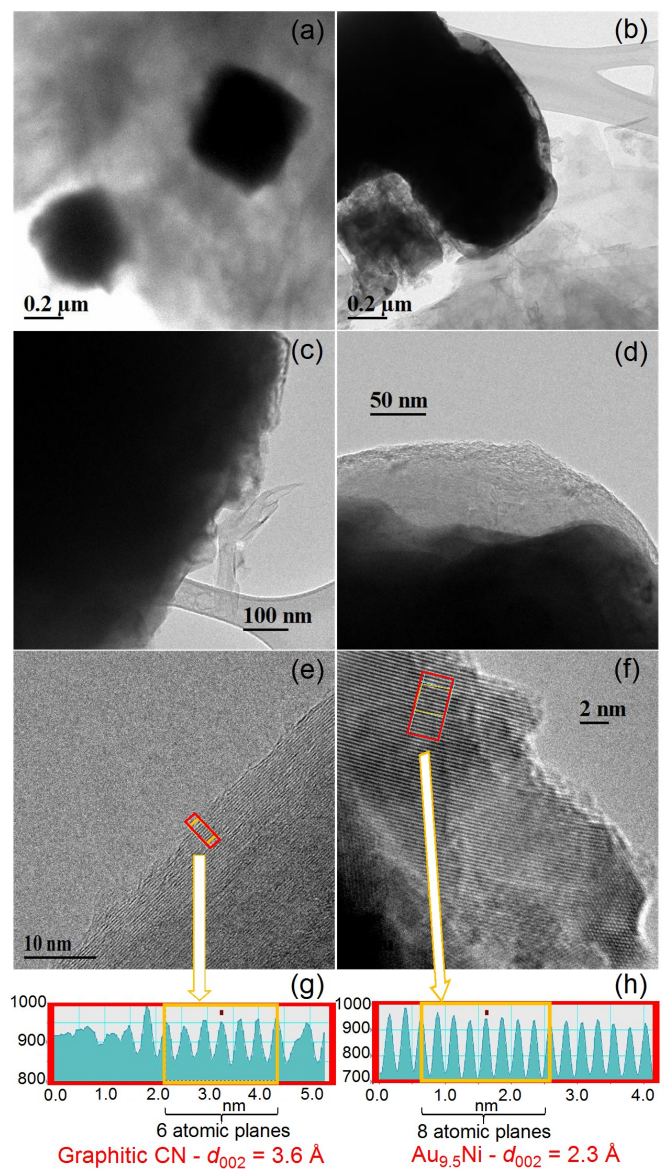

Figure 4. HR-TEM images of AuNi$_{10}$-CN$_l$ 600/Gr (a, c, e, g) and AuNi$_{10}$-CN$_l$ 900/Gr (b, d, f, h) ECs.

The ORR activity and reaction pathway of the *"core-shell"* CN ECs is studied in both acidic (Figure 5a) and alkaline medium (Figure 5b). Results are compared with the performance of the Pt/C ref. in the same conditions. In 0.1 M HClO$_4$ (acidic medium), the ORR onset potential of both AuNi$_{10}$-CN$_l$ 600/Gr and AuNi$_{10}$-CN$_l$ 900/Gr is relatively high, and corresponds to *ca*. 0.53 V and 0.62 V *vs*. RHE, respectively, against 0.95 V *vs*. RHE on the Pt/C ref. With respect to the Pt/C ref., the ring currents of both AuNi$_{10}$-CN$_l$ 600/Gr and AuNi$_{10}$-CN$_l$ 900/Gr are *ca*. 3-4 times larger. These ring currents are attributed to the oxidation of the H$_2$O$_2$ formed as the product of the 2-electron reduction of O$_2$. In the alkaline medium (Figure 5b), the ORR activity on AuNi$_{10}$-CN$_l$ 600/Gr and AuNi$_{10}$-CN$_l$ 900/Gr improves greatly if compared to that determined in acidic solution. The ORR onset potentials are as high as 0.81 V and 0.83 V *vs*. RHE for AuNi$_{10}$-CN$_l$ 600/Gr and AuNi$_{10}$-CN$_l$ 900/Gr, respectively, against a value of 0.95 V *vs*. RHE registered for the Pt/C ref. The lower ORR overpotential reached on the AuNi$_{10}$-CN$_l$ 900/Gr electrocatalyst with respect to AuNi$_{10}$-CN$_l$ 600/Gr is attributed to an enhanced electronic effect due to a synergic contribution of the gold and nickel metal sites to the ORR process (15). The association of nickel with gold most likely modulates the electron configuration of Au d-orbitals, thus improving the adsorption and activation of O$_2$ on Au active sites with respect to the sites found in pristine unalloyed Au (19). Accordingly, the best Au-Ni interactions (and correspondingly, the most pronounced electronic effect, and the lowest ORR overpotential) is revealed as the alloying is more extensive, *i.e.*, for AuNi$_{10}$-CN$_l$ 900/Gr. In the acid environment Ni is also expected to exert a bifunctional effect, thus facilitating the protonation of the ORR intermediates owing to its strong Lewis acid character. This improves the overall ORR kinetics (16).

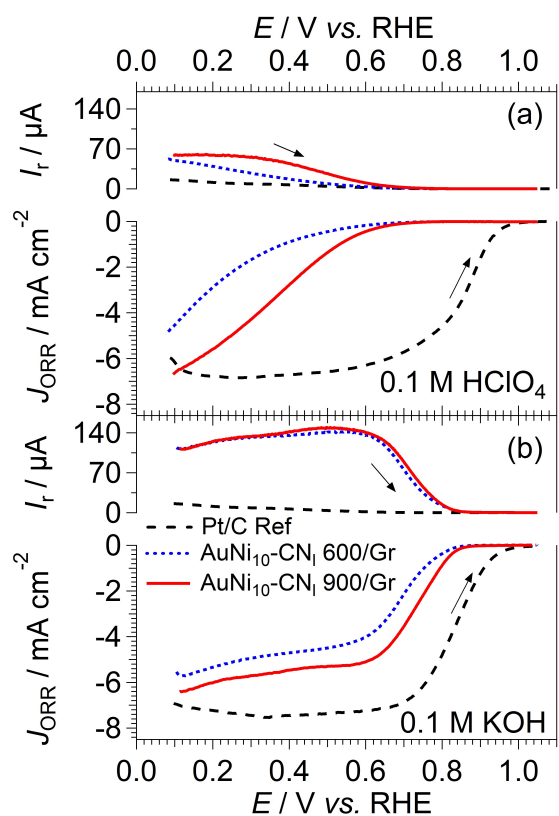

Figure 5. CV-TF-RRDE positive-going sweeps of AuNi$_{10}$-CN$_l$ 600/Gr, AuNi$_{10}$-CN$_l$ 900/Gr and the Pt/C ref. in O$_2$-saturated 0.1 M HClO$_4$ (a) and 0.1 M KOH (b); $v$ = 20 mV·sec$^{-1}$ at $T$ = 25°C.

The amount of generated hydrogen peroxide, shown in Figure 6, illustrates how selective the ECs are towards the 4-electron ORR pathway. In both acidic and alkaline medium, H$_2$O$_2$ production is larger on AuNi$_{10}$-CN$_l$ 600/Gr and AuNi$_{10}$-CN$_l$ 900/Gr than on the Pt/C ref. This low selectivity towards the complete 4-electron ORR into H$_2$O is likely due to the low adsorption strength of the oxygenated reaction intermediates, including H$_2$O$_{2,ad}$, on the electrocatalytically-active Au surface. This phenomenon hinders a further reduction into H$_2$O and favors the release of the intermediates inside the electrolyte as H$_2$O$_2$ (20).

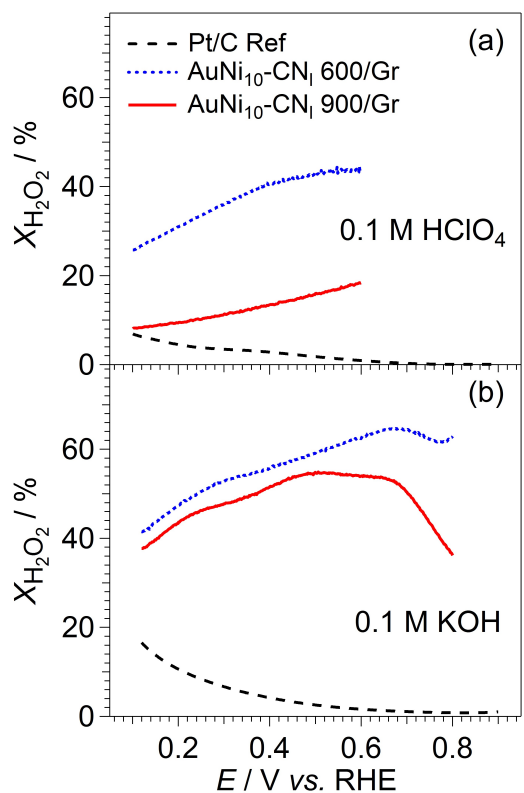

Figure 6. Percentage of H$_2$O$_2$ produced during the ORR as obtained from the CV-TF-RRDE positive-going sweeps (Figure 5) for AuNi$_{10}$-CN$_l$ 600/Gr, AuNi$_{10}$-CN$_l$ 900/Gr and the Pt/C ref. in O$_2$-saturated 0.1 M HClO$_4$ (a) and 0.1 M KOH (b); $v$ = 20 mV·sec$^{-1}$ at $T$ = 25°C.

The Tafel plots reported in Figure 7a illustrate the rather slow ORR kinetics on the two CN-based *"core-shell"* ECs compared to the Pt/C ref. in acidic medium. A low adsorption strength of O$_2$-species could be responsible for the hindrance of the first ORR step, *i.e.* the first electron transfer to O$_{2,ad}$ (21). In alkaline medium, the Tafel slopes of AuNi$_{10}$-CN$_l$ 600/Gr, AuNi$_{10}$-CN$_l$ 900/Gr and the Pt/C ref. are remarkably close to one another, thus indicating that all the ECs share a similar ORR mechanism.

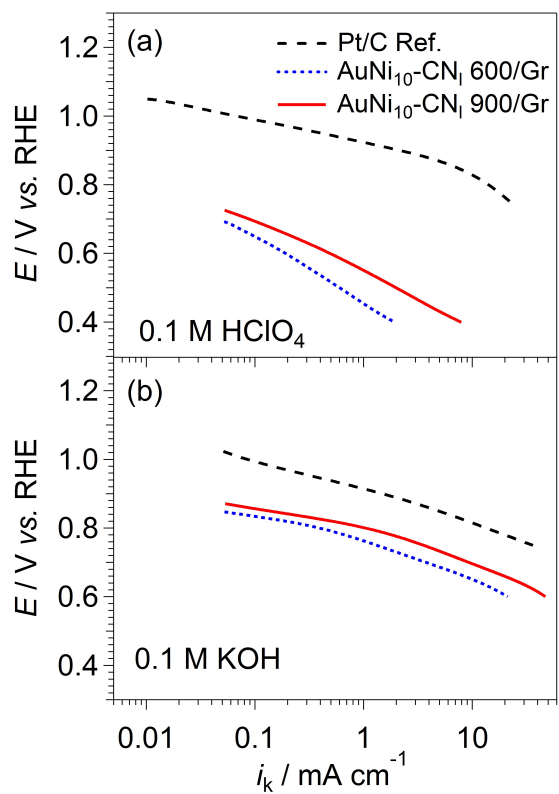

Figure 7. Tafel slopes obtained from the CV-TF-RRDE positive-going sweeps (Figure 5) for AuNi$_{10}$-CN$_l$ 600/Gr, AuNi$_{10}$-CN$_l$ 900/Gr and the Pt/C ref. in O$_2$-saturated 0.1 M HClO$_4$ (a) and 0.1 M KOH (b); ν = 20 mV·sec$^{-1}$ at $T$ = 25°C.

**Conclusions**

In this report, for the first time it was possible to prepare *"core-shell"* ECs based on graphene nanoplatelets as the support *"core"* and a CN matrix embedding AuNi$_x$ alloy NPs as a *"shell"*. These ECs show an efficient ORR performance in an alkaline environment and are obtained by customizing and improving a unique preparation protocol previously devised in our laboratory. Two ECs are synthesized, which are distinguished on the basis of the pyrolysis temperature ($T_f$). Actually, the ECs are pyrolysed at either $T_f$ = 600 or 900°C yielding *"AuNi$_{10}$-CN$_l$ 600/Gr"* and *"AuNi$_{10}$-CN$_l$ 900/Gr"* *"core-shell"* materials, respectively.

The investigation of the chemical composition of the ECs evidences that N atoms are present in the CN *"shell"* matrix covering the *"cores"* of graphene nanoplatelets. In this *"shell"* the Au/Ni molar ratio closely matches the expected nominal values, thus witnessing that the proposed preparation protocol is capable to yield products with a well-controlled chemical composition. The graphitization degree of the CN *"shell"* is enhanced as $T_f$ is raised from 600 to 900°C. In both ECs metallic nanoparticles with well-defined morphologies are revealed, which are supported onto N- and C-based *"coordination nests"* of the CN *"shell"* matrix. The pyrolysis at $T_f$ = 600°C led to materials with a less extensive formation of AuNi$_x$ alloys. Indeed, Ni NPs are clearly evidenced in AuNi$_{10}$-CN$_l$ 600/Gr. In ECs prepared at $T_f$ = 900°C, the formation of alloys is complete, revealing that the AuNi$_x$ NPs exhibit themselves an entire metal *"core-shell"* structure, where a gold-rich Au$_{9.5}$Ni *"shell"* covers a much more abundant AuNi$_{12.3}$ *"core"*.

The here proposed ECs exhibit a remarkable ORR activity in an alkaline environment, showing that AuNi$_{10}$-CN$_l$ 900/Gr exhibits the highest ORR onset potential: 0.83 V *vs.* RHE (for the sake of comparison, the onset potential of the Pt/C. ref in the same conditions is *ca.* 0.95 V *vs.* RHE). It is demonstrated that the ORR performance and selectivity in the 4-electron mechanism improve as $T_f$ is raised from 600 to 900°C. This result is interpreted considering that the ORR process is actually modulated by both electronic and bifunctional effects involving both Au and Ni metal sites. in particular, Ni atoms: (i) facilitate by an electronic effect the adsorption of $O_2$ onto the Au-based ORR active sites; (ii) in the acid environment, they enhance the protonation of the ORR intermediates by a bifunctional effect, promoting the overall ORR kinetics; and (iii) improve the adsorption strength of the $H_2O_2$ intermediate to the Au metal sites, facilitating the complete reduction of $O_2$ to water. These effects are more pronounced in $AuNi_{10}$-$CN_l$ 900/Gr, where the alloying between Au and Ni is very efficient, generating stable alloy NPs with a metallic *"core-shell"* composition. This improves the electronic and bifunctional effects in the ORR process.

More detailed investigations on the interplay between the electrochemical activation processes, the physicochemical properties, the morphology and the electrochemical performance of the ECs in the ORR are still underway and will be the subject of a further contribution. This preliminary study is a first important milestone in the path leading to the development efficient *"core-shell"* CN-based (the *"shell"*) ECs comprising graphene supports (the *"core"*) for application in advanced energy conversion and storage devices such as fuel cells and metal-air batteries.

## Acknowledgements

This project has received funding from the European Union's Horizon 2020 research and innovation programme under grant agreement No. 696656 – GrapheneCore1. This project has also been supported by the Strategic Project of the University of Padova *"From Materials for Membrane-Electrode Assemblies to Electric Energy Conversion and Storage Devices – MAESTRA"*.